\documentclass[sigconf]{acmart}
\AtBeginDocument{%
  }

\copyrightyear{2025}
\acmYear{2025}
\setcopyright{rightsretained}
\acmConference[CHI EA '25]{Extended Abstracts of the CHI Conference on
Human Factors in Computing Systems}{April 26-May 1, 2025}{Yokohama, Japan}
\acmBooktitle{Extended Abstracts of the CHI Conference on Human Factors in
Computing Systems (CHI EA '25), April 26-May 1, 2025, Yokohama,
Japan}
\acmDOI{10.1145/3706599.3720099}
\acmISBN{979-8-4007-1395-8/2025/04}

\begin{document}

\title{Privacy Meets Explainability: Managing Confidential Data and Transparency Policies in LLM-Empowered Science}

\author{Yashothara Shanmugarasa}
\email{yashothara.shanmugarasa@data61.csiro.au}
\orcid{0000-0002-6414-9416}
\authornote{Corresponding Author}
\affiliation{%
  \institution{CSIRO's Data61}
  \city{Sydney}
  \state{NSW}
  \country{Australia}
}

\author{Shidong Pan}
\authornote{The majority of this work was carried out while the author was affiliated with Data61, CSIRO, Australia, with support from a Data61 scholarship.}
\email{Shidong.Pan@anu.edu.au}
\orcid{0000-0002-2162-0407}
\affiliation{%
  \institution{School of Computing, Australian National University}
  \city{Canberra}
  \state{ACT}
  \country{Australia}
}

\author{Ming Ding}
\email{ming.ding@data61.csiro.au}
\orcid{0000-0002-3690-0321}
\affiliation{%
  \institution{CSIRO's Data61}
  \city{Sydney}
  \state{NSW}
  \country{Australia}
}

\author{Dehai Zhao}
\email{dehai.zhao@data61.csiro.au}
\orcid{0000-0003-3637-4939}
\affiliation{%
  \institution{CSIRO's Data61}
  \city{Sydney}
  \state{NSW}
  \country{Australia}
}

\author{Thierry Rakotoarivelo}
\email{thierry.rakotoarivelo@data61.csiro.au}
\orcid{0000-0001-7698-6214}
\affiliation{%
  \institution{CSIRO's Data61}
  \city{Sydney}
  \state{NSW}
  \country{Australia}
}




\renewcommand{\shortauthors}{Yashothara et al.}

\begin{abstract}
  
As Large Language Models (LLMs) become integral to scientific workflows, 
concerns over the confidentiality and ethical handling of confidential data have emerged. 
This paper explores data exposure risks through LLM-powered scientific tools,
which can inadvertently leak confidential information, 
including intellectual property and proprietary data, 
from scientists' perspectives. 
We propose ``DataShield", 
a framework designed to detect confidential data leaks, 
summarize privacy policies, 
and visualize data flow, 
ensuring alignment with organizational policies and procedures. 
Our approach aims to inform scientists about data handling practices, 
enabling them to make informed decisions and protect sensitive information. 
Ongoing user studies with scientists are underway to evaluate the framework's usability, trustworthiness, and effectiveness in tackling real-world privacy challenges.

\end{abstract}




\maketitle

\section{Introduction}

Artificial Intelligence (AI) represents a transformative technology, revolutionizing various industries by automating complex tasks and providing intelligent solutions. 
A prominent development in this field is the rise of pre-trained Large Language Models (LLMs), 
which have set new standards in natural language processing (NLP), enabling machines to generate human-like text and perform sophisticated tasks with remarkable accuracy \cite{dyde2023documentation}. 
LLMs have become integral to scientific workflows, empowering researchers to efficiently query, analyze, and synthesize vast datasets using LLM-powered systems \cite{brown2020language}. 
The 2024 Nobel Prizes in Biology and Physics, awarded for advancements in machine learning, highlight the growing role of AI and LLMs in driving scientific innovation \cite{nobel2024}. 
In particular, 
the development of LLM agents capable of accessing external tools, 
making autonomous decisions, and performing multiple functions, 
has further enhanced the application of AI in scientific research.

However, scientists frequently work with confidential and intellectual property (IP) data belonging to their organization or individuals, 
which raises significant privacy concerns. 
These concerns stem from using LLM applications that may share sensitive information with service providers, 
often without a clear understanding of the associated risks, 
company policies on IP data protection, 
or the potential involvement of third-party tools. 
LLMs have increased data exposure compared to traditional software applications (typically require specific predefined fields for service access) as LLM prompts can include unrestricted information, 
and their conversational nature often encourages users to disclose more than originally intended \cite{10.1145/3613904.3642385}. 
LLM systems are fed diverse types of information from various sources in their prompts, 
potentially revealing more contextual data beyond the direct sensitive data in the prompts. 
For these reasons, 
many tech companies have implemented restrictions on LLM tools. 
For example, Samsung banned the use of tools like ChatGPT following an internal data incident \cite{forbesSamsungBans}.

Recently, the growing regulatory focus on personal data, particularly Personally Identifiable Information (PII), 
through frameworks such as GDPR and guidelines from organizations like the National Institute of Standards and Technology (NIST), 
along with decisions by government agencies like the US National Science Foundation and the Italian government \cite{nistNationalInstitute, nsf}, 
has spurred research efforts in privacy and confidentiality preservation \cite{lin2024promptcrypt, microsoftHomeMicrosoft, lakeraIntroducingLakera}. 
These studies aim to protect personal data within prompts by identifying and highlighting PII or encrypting them in various formats. 
However, while these efforts primarily focus on detecting and protecting PII, 
leaking confidential data extends far beyond PII when using LLMs in scientific contexts. 
This includes proprietary and IP data critical to organizations, 
such as gene or protein sequences, material names, chemical formulations, and algorithms. 
These types of data are highly sensitive for organizations but do not fall under the PII category, 
a gap that remains largely unaddressed in existing research. 
Moreover, LLM agents enable seamless integration with multiple external tools, 
allowing users to perform various tasks effortlessly through a single prompt. 
However, this convenience also increases the vulnerability of LLM platforms, 
as many users (e.g., scientists) remain unaware that their data traverses multiple pathways beyond the LLM service providers alone. 
This aspect of data vulnerability has not been adequately addressed in existing literature, particularly the need to inform users about the involvement of external tools, 
their privacy policies, 
and their compliance with the organization's internal policies.

To address the privacy and ethical concerns of using LLM-empowered AI tools, 
we propose ``DataShield – Explainable Shield for Confidential Data," framework.
Our approach includes three key components: 
i) a confidential data detection module, 
ii) a policy summarization module that creates clear ``privacy nutrient labels" based on the privacy policies of external tools in LLM-empowered systems while also summarizing the organization's internal policies, and 
iii) a visualization dashboard to display data flow, triggered actions, and recommendations.
Our research goal is to inform users better, 
particularly research scientists handling confidential information about the potential risks to their data when interacting with LLM-based tools. 
Our approach identifies what types of confidential data may be inadvertently exposed and whether such exposure violates internal company policies, 
such as the code of conduct, 
especially when third-party external tools are involved. 
The code of conduct outlines the dos and don'ts regarding data handling, confidentiality, and privacy.
By detecting potential data leaks and analyzing relevant privacy policies, 
our framework ensures that users are fully aware of how their data is handled and empowered to make informed decisions.

\begin{figure*}[!ht]
    \centering
   \includegraphics[width=1\linewidth]{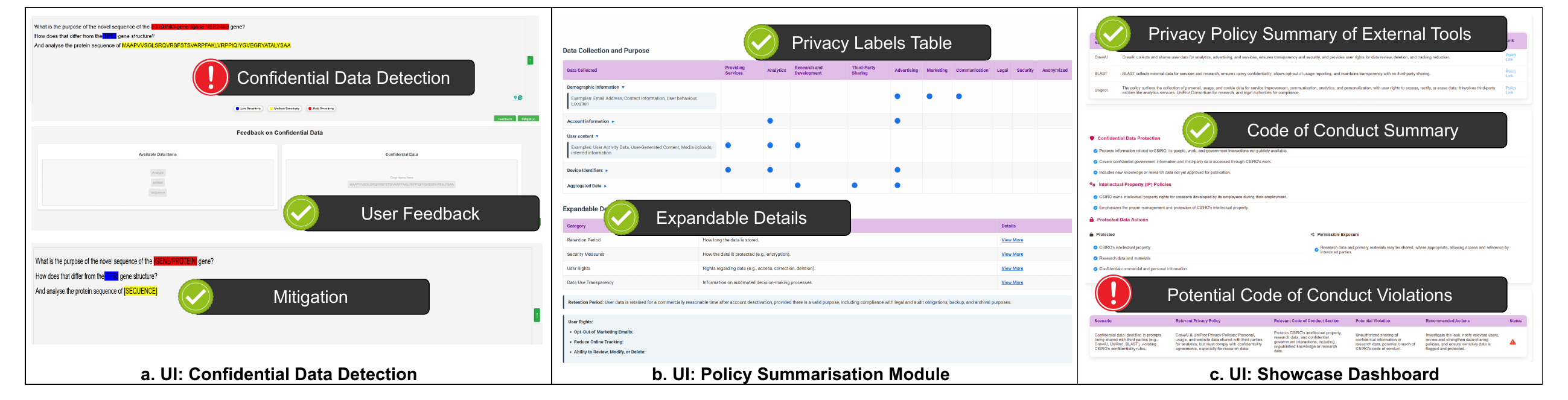}
    \caption{The overview of ``DataSheild" framework with three modules output: User Interface: Clear version can be found \href{https://drive.google.com/file/d/1FRNvTNWgI-Jnwr4Xj8Mo3KGTu-NoH7gy/view?usp=sharing}{here}}
    \label{fig:teaser}
    \Description{The overview of ``DataSheild" framework with three modules: A Confidential Data Detection Module, A Policy Summarization Module, and A Visualization Dashboard Module. It shows the user interface of the ``DataSheild" framework, highlighting the detailed features of each module.}

\end{figure*}

We plan to conduct user studies to evaluate scientists' perceptions of the desirability, trustworthiness, and suitability of our proposed framework in addressing their privacy and ethical concerns when using LLM systems. 
Notably, we will use synthetic data as the hypothetical ``confidential" data and open-sourced localized LLMs (e.g., Claude3-Opus) to guarantee the integrity of our user study.

\subsection{Research Questions}

The primary research question guiding this study is:  
\textit{How can LLM-powered AI tools for scientific discovery (particularly in genomics) be designed to enhance confidentiality, ethical compliance, and ensuring alignment with organizational standards and minimizing confidential data leakage?}
To address this, we propose a framework with multiple components, each investigating the following sub-research questions:

\begin{enumerate}
    \item \textbf{Confidential Data Exposure in LLM Interactions:}  
    Scientists interacting with LLM-powered AI tools may inadvertently disclose confidential data, such as gene and protein names, without realizing potential risks. Existing AI systems lack mechanisms to detect and mitigate such leaks while maintaining a seamless workflow.  

    \begin{itemize}
         \item \textbf{RQ1:} How can we systematically identify and classify confidential data (e.g., gene and protein names) in scientist-LLM interactions while assessing its sensitivity to provide adaptive risk-based alerts?  
         \item \textbf{RQ2:} How can integrating user feedback and domain-specific contextual knowledge enhance the accuracy and adaptability of confidential data detection?
    \end{itemize}
    
    \textbf{Our Solution:}  
    Our framework incorporates a `confidential data detection' module that automatically identifies and categorizes sensitive terms in scientist-LLM interactions.

    \item \textbf{Lack of Policy Awareness and Compliance Support:}  
    Scientists often struggle with navigating internal and external privacy policies when using multiple LLM-powered AI tools.

    \begin{itemize}
        \item \textbf{RQ3:} How can we effectively summarize and present internal compliance policies and external tools' privacy policies (e.g., genome sequencing tools) to scientists for improved awareness and decision-making?
        \item \textbf{RQ4:} What interactive mechanisms can align external privacy policies with organizational internal policies to support users in proactively managing data-sharing risks?
    \end{itemize}

    \textbf{Our Solution:}  
    Our `policy summarization' module extracts and summarizes key policy elements.

    \item \textbf{Limited Transparency in AI-Driven Data Processing:}  
    The lack of visibility into how LLM systems process user data, trigger actions, and involve third-party tools reduces trust and usability. 

    \begin{itemize}
        \item \textbf{RQ5:} What visualization techniques can enhance user comprehension of data flow, triggered actions, and third-party tool interactions within LLM-powered systems?
        \item \textbf{RQ6:} How does interactive visualization impact scientists' trust, usability perception, and decision-making regarding data confidentiality?
    \end{itemize}

    \textbf{Our Solution:}  
    We introduce a `visualization dashboard' that provides a comprehensive view of data flow, AI-triggered actions, and third-party tool interactions.

    \item \textbf{Usability-Centered Evaluation of Privacy and Explainable AI Frameworks:}  
    Current privacy-preserving AI frameworks lack systematic evaluation methods that assess usability, compliance effectiveness, and trust.

    \begin{itemize}
        \item \textbf{RQ7:} How do scientists perceive the usability, trustworthiness, and desirability of the proposed ``DataShield" framework in mitigating privacy risks?
        \item \textbf{RQ8:} What qualitative and quantitative evaluation methods can assess the effectiveness of the ``DataShield" framework in balancing usability, compliance, and confidentiality concerns?
    \end{itemize}

    \textbf{Our Solution:}  
    We conduct a mixed-method evaluation, combining a quantitative assessment of each component with user studies involving scientists.
\end{enumerate}

Figure \ref{fig:teaser} serves as a teaser image, presenting the user interface (UI) representation of our ``DataShield" framework and providing an engaging overview of our contributions, which are elaborated in the subsequent sections of the paper.

\section{Related Works}

\subsection{Studies on Personally Identifiable Information Detection}

Several research efforts and commercial products are available as plugins to identify PII data in user prompts when interacting with LLM-empowered applications. PII detection methods typically use Named Entity Recognition (NER), predefined policies, LLMs, regular expressions, rule-based logic, or checksums to identify and remove sensitive details across multiple languages and contexts \cite{lin2024promptcrypt, chen2023hide, li2024human, stracSecureSensitive, lakeraIntroducingLakera, microsoftHomeMicrosoft}.
After detecting PII data, mitigation strategies often replace sensitive information with placeholders. For example, EmojiCrypt \cite{lin2024promptcrypt} encrypts data using emojis and math operations, while others substitute or mask data \cite{chen2023hide}. Secure prompt templates and redaction tools also help minimize data leakage \cite{stracSecureSensitive}. \citet{hartmann2024can} suggested obfuscating sensitive queries with high-level descriptions, new problems, or placeholders. TextObfuscator \cite{zhou2023textobfuscator} similarly obfuscates sensitive words while retaining functional meaning.

However, existing approaches focus on PII data, leveraging advanced NER taggers (e.g., spaCy NER \cite{spacyEntityRecognizerSpaCy})
and LLMs trained on PII datasets, making personal data detection relatively straightforward. However, detecting confidential data, such as gene names in genomics or chemical names in manufacturing, is more complex and underexplored. Additionally, high-risk scenarios where LLMs infer sensitive information from seemingly insignificant data are often overlooked \cite{kroger2022personal}. Our framework addresses these gaps by focusing on confidential data detection and indirect data exposure.

\subsection {Studies on Policy Summarization and Privacy Nutrition Labels}


Privacy policies are legally binding documents for organizations to disclose data collection practices, mandated by privacy laws \cite{cui2023poligraph}. However, their complexity and length make them difficult for users to comprehend \cite{mcdonald2008cost}.
This challenge has driven extensive research efforts in policy analysis and summarization, particularly for web and mobile applications.
Early studies, such as OPP-115 \cite{wilson2016creation}, focused on taxonomy creation and manual annotation of privacy policies, while later datasets like \citet{bui2021automated} introduced large-scale corpora for automated analysis. With advancements in NLP, AI-based tools such as Privee \cite{zimmeck2014privee}, Polisis \cite{harkous2018polisis}, PolicyLint \cite{andow2019policylint}, PurPliance \cite{bui2021consistency}, and OVRseen \cite{trimananda2022ovrseen} have emerged using classifiers to extract data types and entities in the privacy policies. PolicyGPT \cite{tang2023policygptautomatedanalysisprivacy}, based on LLMs, categorizes clauses into predefined classes. However, these tools often lack connections between extracted entities, such as which data type is collected, by whom, and for what purpose.
PoliGraph \cite{cui2023poligraph} addressed this by using a knowledge graph to capture relationships between data types, entities, and purposes but produced outputs too complex for users. 
Our two-layer approach builds on PoliGraph, combining its accuracy with LLM-based summarization to produce concise, user-friendly outputs. This method avoids hallucinations and enhances readability compared to standalone LLMs.

\citet{kelley2009nutrition} introduced privacy labels inspired by nutrition labels to simplify privacy policy presentation, offering users a clear overview of data collection, usage, and sharing practices. 
\citet{pan2023toward} proposed a framework for generating privacy nutrition labels from applications' privacy policies.
We also generate privacy nutrition labels from policies but extend the approach by mapping relationships between entities and purposes using PoliGraph to improve user understanding of data collection reasons. Unlike previous approaches that focus on a single policy, our method summarizes privacy policies across multiple tools to create an overall privacy label. We also analyze the company's internal policies and their alignment with external tools' data collection behaviour in their privacy policies, ensuring users can safely use LLMs while maintaining compliance.

\subsection {Studies on Privacy Dashboards}

Privacy dashboards can provide access to personal data in a structured and interactive manner \cite{bier2016privacyinsight}. Server-side tools for privacy settings \cite{googleSignGoogle, primeprojectPRIMEPrivacy}
dominate the market by providing a dashboard that allows users to handle privacy settings. Various approaches aim to raise privacy awareness by visualizing data collection, including privacy dashboards in \cite{kolter2010visualizing} and \cite{kani2012increasing}, which use transaction logs and maps to highlight data by category and purpose. Provenance-based tools \cite{aldeco2008provenance, pulls2013distributed, angulo2015usable,bier2016privacyinsight} visualize data flows and sharing details. 
Other toolkits, such as \cite{angulo2015usable, schufrin2020visualization, raschke2018designing, grunewald2021tilt}, 
align with GDPR transparency principles, 
offering more automated and adaptive use of transparency information. 
However, we found no existing approach that combines all involved external tools in an LLM-powered platform for privacy information visualization.

\subsection{Studies on Explainability and Human Involvement in Responsible LLMs}

To address the gap in model-centered research that lacks user perspectives, recent studies have explored explainability and human involvement to better understand key privacy risks and how existing research meets user needs. \citet{10.1145/3613904.3642385} conducted a human-centered study on user disclosure behaviors and risk perceptions in LLM-based conversational agents. By analyzing sensitive disclosures in ChatGPT conversations and interviewing 19 users, they found that users face trade-offs between privacy, utility, and convenience. However, users' erroneous mental models and dark patterns in system design limited privacy awareness, while human-like interactions encouraged more sensitive disclosures, complicating these trade-offs.
Similarly, studies \cite{10.1145/3613905.3636301, liao2023ai, kapania2024m, barman2024beyond} examine the responsible integration of LLMs into research workflows. They outline a research agenda on using LLMs as research tools, addressing open empirical and ethical evaluation questions. These works explore transparency and responsible AI, focusing on issues such as perceived lack of control, distributed responsibility in the LLM supply chain, conditional ethical engagement, and competing priorities.

There are some studies \cite{shruti2024responsible, saba2023towards} explore black-box models and their decision-making processes, which fall outside our scope; so we do not delve further into this area. As noted in \cite{ehsan2024human, kandhari2024responsible}, algorithmic transparency alone is insufficient for AI explainability, which extends beyond merely "opening" the black box. We address explainability by ensuring transparency in confidential data involvement, tool interactions, and policy alignment. While most studies identify requirements or present position papers through user studies, we found no other work combining technical contributions with user studies to evaluate them.

\subsection{Our Unique Contributions}

This paper makes a unique contribution by focusing on the privacy, confidentiality, and explainability aspects of LLM-empowered AI systems from the perspective of research scientists and organizations. 
We are the first study to address three key components to enhance the safety, ethical management, and explainability of these systems, 
to prevent confidential data leakage and to improve user trust and decision-making. 
Our approach integrates LLM-based automation, a human-in-the-loop model, and extends privacy considerations to confidentiality. We also conduct user studies with research scientists on confidentiality and compare our work to existing approaches.

Within these components, our contributions are distinctive. 
For confidential data detection, 
we focus on sensitive data like gene and protein names, 
which presents challenges that prevent us from relying solely on existing NER taggers or LLMs. 
In the second component, 
we use a two-layer summarization approach to maintain entity connections while ensuring readability, 
and our method differs from others by aligning external tool usage with the organization's internal policies through the summarization of its code of conduct.
The third component features a dashboard to provide users with an overview of privacy, displaying detected confidential data, privacy policies, and compliance with the organization's internal codes of conduct.
Furthermore, 
we conduct user studies with scientists working with such data, 
gathering their feedback to evaluate the desirability, trustworthiness, and suitability of our approach in addressing privacy and ethical concerns in LLM systems.

\section{Proposed solution}

\begin{figure*}[!ht]
    \centering
    \includegraphics[width=1\textwidth, height=\textheight, keepaspectratio, alt={Workflow Diagram of DataShield}]{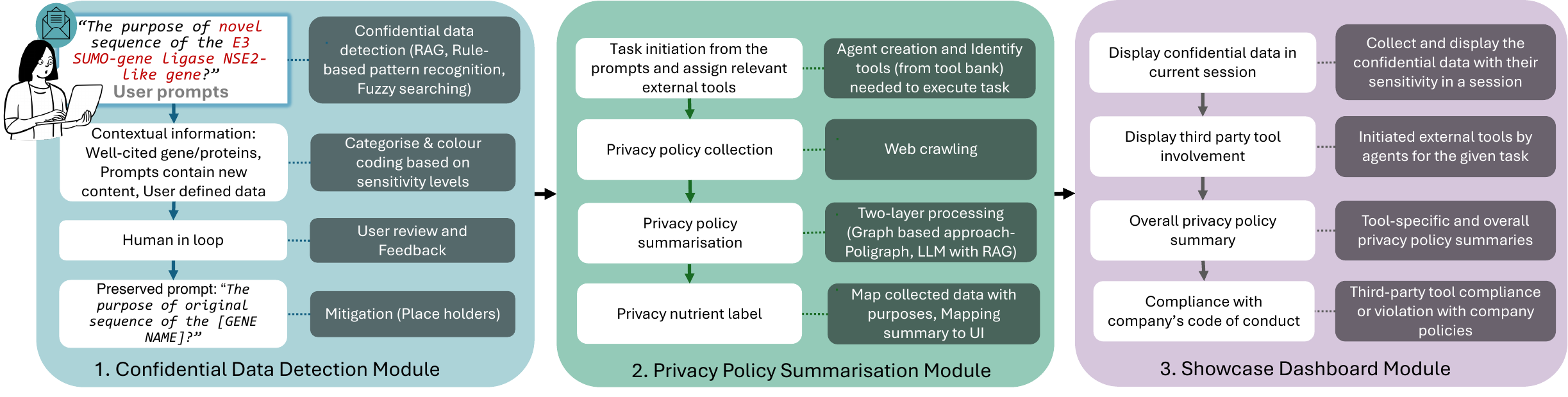}
    \caption{Workflow Diagram of ``DataShield" – Explainable Shield for Confidential Data (RAG: Retrieval-Augmented Generation)}
    \label{fig:workflow}
    \Description{Workflow Diagram of ``DataShield" – Explainable Shield for Confidential Data. It illustrates the process from the moment a user enters confidential data in a prompt to the final display of the overall workflow and policies involved in the session between the user and the LLM-powered AI system. Additionally, it highlights the techniques used at each step to achieve this process.}
\end{figure*}

We propose ``DataShield – Explainable Shield for Confidential Data," 
a comprehensive framework that includes a confidential data detection module (to identify confidential information), 
a policy summarization module (to condense complex privacy policies of external tools into a clear ``privacy nutrient label" and summarize the code of conduct of companies), 
and a visual dashboard (displaying data flow, triggered actions, and recommendations), 
especially aimed at scientists. 
The workflow diagram of the overall framework is shown in Figure \ref{fig:workflow}, 
consisting of three main modules and a user study component to evaluate our system.

\subsection{A Confidential Data Detection Module}

The ``DataShield" process begins with the confidential data detection module (RQ1, RQ2), 
which analyzes user prompts to identify potentially confidential data (currently focused on gene and protein data). 
For example, consider the following prompt provided by a biologist: ``What is the purpose of the novel sequence of the \textit{E3 SUMO-gene ligase NSE2-like gene}? And analyze this protein sequence `\textit{MAAPVVSGLSRQVRSFSTSVARPFAKLVRPPIQIYGVEGRYATALYSAA}'." 
This prompt contains confidential information, 
such as the gene name (E3 SUMO-gene ligase NSE2-like gene) and the protein sequence. 
The module is designed to detect and flag such confidential data when entered into the system (Figure \ref{fig:teaser}: a).

Recognizing the complexities of natural language and the potential for variations in phrasing and terminology, 
the module incorporated three complementary techniques. 
Firstly, rule-based pattern recognition uses predefined rules to detect specific phrases and data formats, 
identifying confidential protein sequence data based on its pattern. 
To enhance the module’s ability to handle more nuanced language and contextual information, 
a Retrieval Augmented Generation (RAG) approach was integrated. 
This approach involved utilizing LLM to extract relevant information from a knowledge base containing a collection of public gene and protein names sourced from the Uniprot library \cite{uniprot2015uniprot}.
This enabled the system to generate contextually relevant responses and enhance the accuracy of identifying confidential data by incorporating gene and protein names from the knowledge base, 
providing more accurate information than relying solely on the LLM, as it learns from these examples. 
Finally, we incorporated a human-in-the-loop approach to define additional confidential data as specified by the company or individual. 
The system then checks these user-defined data against prompts using fuzzy logic searching, 
enabling it to identify potential matches even with minor discrepancies.

In addition to detecting confidential data, 
we incorporate contextual information, 
such as well-cited genes or proteins, 
novelty exposure in the prompt, 
and user-defined confidential information, 
to categorize the severity of the data. 
The colour-code scheme is as follows: 
red for the high sensitivity, 
yellow for the medium sensitivity, 
and blue for the low sensitivity. This design choice is based on its intuitive and widely recognized use in risk assessment \cite{jensen2022risk}, allowing scientists to gauge the level of confidentiality associated with their input immediately.

Additionally, 
we have addressed indirect data exposure through prompts, 
where the inference capabilities of LLMs can reveal sensitive information that is not explicitly mentioned in the prompt. For instance, in the prompt, "We have identified Gene\_B in a wild maize relative, sharing a conserved domain with an Arabidopsis receptor involved in salt stress signalling. How can we annotate and validate its role in salinity tolerance?", although Gene\_B is unnamed, references to salt stress and conserved domains imply potential candidates like SOS1 or related SOS pathway components, highlighting the risk of indirect data exposure.

The next feature of this module allows users to provide feedback on the detected confidential data, 
enabling them to mark it as confidential or not, 
with the system learning from this feedback. 
Furthermore, 
we propose a simple mitigation technique using placeholders (e.g., [GENE\_NAME]) to redact confidential data from the prompt.

\subsection{A Policy Summarization Module}

The second core component of ``DataShield"  is a Policy Summarization Module (RQ3, RQ4), which aims to provide users with clear and concise summaries of complex privacy policies from external tools and internal company policies, 
such as the code of conduct (i.e., outlining what scientists can do to comply with company guidelines), thereby making it easier for them to understand the implications of sharing their data. 
The process begins by identifying relevant external tools from the tool bank, including 40 CrewAI tools \cite{crewaiToolsCrewAI},
LangChain \cite{langchainTools},
and gene-related tools \cite{brouard2019gatk}.
These tools were particularly aligned with the genomics use case scenario we focused on. 
To identify the potential tools, 
we created an agentic environment based on the user prompt, 
which helped us identify the most appropriate tools for executing tasks specified by the users. 
After determining the relevant tools, 
we then scraped the privacy policies associated with each of these tools online. 
The privacy policies collected for each tool are then processed using a two-layer summarization approach. 
The first layer employed a graph-based method, PoliGraph \cite{cui2023poligraph}, to analyze the structure and extract privacy practice disclosures from privacy policies. 
This approach mapped relationships between identified data entities, data types, collection purposes, and policy procedures into a structured format. 
However, the output of PoliGraph, 
while comprehensive, 
is not in a human-readable format and is often lengthy. 
A second layer was introduced to further process the extracted information by 
utilizing an LLM with the RAG strategy. 
This layer complemented the graph-based analysis by extracting important policy information, 
identifying important clauses and conditions, 
and generating concise, meaningful summaries that highlighted essential aspects of the policies. 
To enhance the readability further, 
we transformed the extracted content in the form of ``privacy nutrient labels”. 
The labels provide a concise summary of the most critical elements of the privacy policy, 
including data types, purposes, retention periods, security measures, user rights, and third-party involvement. 
By presenting this information in an intuitive format (as shown in Figure \ref{fig:teaser}: b), 
users could quickly understand the key implications of each policy, 
promoting transparency.

Similarly, 
we leveraged another LLM empowered by RAG to summarize the company’s internal policies, 
resulting in insights such as Confidential Data, IP Policies, Protected vs. Exposed Information, Violations of Confidentiality and IP data, and additional policies required for compliance. 
Using the LLM, we assessed the summarized privacy policies of external tools and the company’s code of conduct to scrutinize compliance and potential violations (Figure~\ref{fig:teaser}: c).

\subsection{A Visualization Dashboard Module}

The Visualization Dashboard Module (RQ5, RQ6) (Figure \ref{fig:teaser}: c), 
served as the central interface for presenting the results of the confidential data detection and summarization processes with a clear and intuitive view.
The dashboard displays the confidential data detected 
highlighting its sensitivity level. 
It also provides information about any relevant external tools involved in the prompt execution, along with individual tool privacy nutrient labels. 
Critically, the dashboard displayed the overall privacy policy summaries generated by the Policy Summarization Module, 
along with their compliance/violation with summarized internal company policies, 
providing users with a comprehensive understanding of the applicable privacy practices and compliance requirements. 
This integration of information within a single dashboard aimed to empower users with the knowledge they needed to make informed decisions about their data.

\subsection{User Study}

A user study (RQ7, RQ8) will be conducted to evaluate the effectiveness and usability of the integrated framework by interacting with it. 
We obtained ethical approval from our Human Research Ethics Committee.
Participants in this study would be composed by scientists from various subjects, 
such as genomics, computer science, and chemical and material engineering, 
who utilize LLM-empowered AI tools in organizations. 
To expand beyond genomics, 
we will enhance the confidential data detection module to accommodate confidential data from other domains. 
We anticipate 30 to 40 scientists will participate in the user study. 
Participants will receive a link to the ``DataShield" framework, a questionnaire, and a consent form with privacy information. The questionnaire is designed to ensure that no personally identifiable information is collected. The study consists of three sections:
1) demographic questions assessing familiarity with LLM-empowered tools and privacy concerns; 
2) hands-on experience with ``DataShield" using a synthetic dataset (will be given to the users) containing prompts that a scientist might ask; and 
3) a user experience survey on how ``DataShield" addresses user concerns. The synthetic dataset, created with ChatGPT-4o using publicly known gene or material names and synthetic company names, contains no real or confidential data, ensuring no privacy risks. We use a prompt in ChatGPT to create the synthetic dataset.
This approach ensures the integrity of the synthetic dataset while focusing on testing privacy management mechanisms effectively. 
Humans will review every prompt ChatGPT generates to ensure it is produced as expected and does not unintentionally leak any confidential information. 
To enhance participant security, 
they will interact with DataShield, 
powered by Claude3-Opus localized, 
open-source LLM—instead of an online LLM application. 
Claude3-Opus serves as the backend for the DataShield framework, 
offering a secure and flexible environment while ensuring that data privacy is maintained throughout the testing process.
Participants are asked to use the ``DataShield'' to simulate their daily usage scenario for 30 minutes. 

Following the hands-on experience, 
participants will be surveyed on several key aspects, 
including the desirability of the system, 
the level of trust users placed in its accuracy, 
the perceived information load, 
the accuracy of the information provided, overall user satisfaction, 
and suggestions for improvement. 
The data gathered from this user study provided valuable insights into the system's strengths and weaknesses, 
informing further refinements and improvements. 



\section{Preliminary Results}

This research utilizes a mixed-methods approach, combining qualitative methods (through user studies) and quantitative techniques to evaluate the effectiveness of our framework, both overall and at the level of individual modules.

The effectiveness of the confidential data detection module was quantitatively assessed using accuracy by comparing it with baseline biomedical NER tools such as BERN2 \cite{sung2022bern2}, HunFlair1 \cite{weber2021hunflair}, and HunFlair2 \cite{sanger2024hunflair2}, as well as various LLMs, including GPT-4o, and local LLMs such as Mistral-Large-2407, Claude3-Opus, Claude3-Sonnet, Llama-3.1-70b, and Claude2. 
We used 500 test sentences from the BC2GM dataset \cite{smith2008overview}, 
which was used to train the BERN2 model. 
The results, summarized in Table \ref{tab:results_cdd}, 
show that while GPT-4o excelled in accuracy, 
its precision and F1 scores were lower. 
The BERN2 model yielded strong performance; 
However, these results may not be generalized well to other scenarios, 
as the test dataset was the same one used to train the BERN2 model. 
Future work will evaluate performance across additional public datasets to ensure generalizability across diverse models and scenarios. 
For the user study experiments, 
we selected Claude3-Opus due to safety and privacy considerations. 
Although participants are advised to interact with synthetic datasets during testing, 
Claude3-Opus was chosen as an additional precaution to mitigate risks in the event of inadvertent disclosure of confidential data. 
Claude3-Opus achieved reasonable accuracy, precision, recall, and F1 scores, balancing safety and effectiveness. Moreover, organizations can avoid relying on external LLM service providers to detect confidential data by deploying a small local LLM on their premises. 
This approach enhances confidentiality in handling highly sensitive data and reduces maintenance costs by utilizing a compact local model to process data before sharing them with external providers.

\begin{table}[h!]
\centering
\caption{Performance comparison of various tools for confidential data detection. (RAG: Retrieval Augmented Generation)}
\small
\setlength{\tabcolsep}{3pt}
\resizebox{\columnwidth}{!}{%
\begin{tabular}{@{}lcccc@{}}
\toprule
\textbf{Tool} & \textbf{Accuracy (\%)} & \textbf{Precision (\%)} & \textbf{Recall (\%)} & \textbf{F1 Score (\%)} \\ \midrule
BERN2              & 90.01  & 79.06  & 90.01  & 84.18  \\
HunFlair1          & 74.19  & 93.80  & 74.19  & 82.85  \\
HunFlair2          & 70.38  & 91.31  & 70.38  & 79.49  \\
GPT-4o + RAG            & 97.56  & 69.78  & 97.56  & 81.36  \\
Mistral-Large-2407 + RAG & 75.08  & 90.19  & 75.15  & 81.98  \\
Claude3-Opus + RAG      & 76.73  & 94.10  & 76.80  & 84.57  \\
Claude3-Sonnet + RAG     & 65.59  & 92.73  & 65.59  & 76.83  \\
Llama-3.1-70b + RAG     & 68.29  & 89.85  & 68.29  & 77.60  \\
Claude2 + RAG           & 47.99  & 84.75  & 47.99  & 61.28  \\ \bottomrule
\end{tabular}%
}
\label{tab:results_cdd}
\end{table}


The outputs of the Policy Summarization Module and the Visualization Dashboard are illustrated in Figure \ref{fig:teaser}: c,d. 
The Policy Summarization Module will be quantitatively evaluated using a question-answer approach by deriving a set of questions related to the policy, 
such as the purpose of data collection and retention periods. 
Two reviewers 
will assess the entire policy and its summary to determine how often the summarization provides correct answers based on their agreement.
The Visualization Dashboard will be evaluated through a user study,
which is currently underway. 

\section{Discussion and Future Work}

This paper addresses the confidentiality and ethical management challenges of LLM-powered AI tools for scientific discovery from the perspective of scientists. In future work, We will evaluate our approach through user studies with scientists, combining qualitative insights and quantitative assessments. The confidential data detection module will be assessed for accuracy across data domains, and policy summarization will be tested using a question-answer approach and LLM-based evaluations. We also aim to extend the framework to material and data science for broader applicability. Future work will also enhance system reliability by mitigating hallucinations using strategies like prompt engineering and reinforcement learning from human feedback. This aligns with our human-in-the-loop framework, ensuring real-time feedback and protection of sensitive data, thereby promoting the safe adoption of LLMs in scientific research. We will also explore DataShield’s scalability in multi-user environments to evaluate its performance in collaborative research contexts.

\bibliographystyle{ACM-Reference-Format}
\bibliography{sample-base}

\end{document}